\documentclass[11pt]{article}
\usepackage{times}
\usepackage{sidecap}
\usepackage{geometry}
\usepackage{floatrow}
\usepackage{float}
\usepackage{hyperref}
\usepackage{graphics,graphicx}
\usepackage{subcaption}
\usepackage{epsf,wrapfig}
\usepackage{tablefootnote}
\usepackage{natbib,bm}

\usepackage{paralist}
\let\olditem\item
\renewenvironment{thebibliography}[1]{%
  \section*{\refname}
  \let\par\relax
  \renewcommand{\item}[1][]{\olditem[\textbullet]}%
  \inparaenum}{\endinparaenum}

\newcounter{aff}

\DeclareGraphicsExtensions{.jpg,.pdf,.png,.eps}

\usepackage{caption}
\usepackage{xcolor}

\usepackage{tcolorbox}

\bibpunct{(}{)}{;}{a}{}{,}

\usepackage{soul}

\definecolor{darkcyan}{rgb}{0.0, 0.0, 0.7}

\definecolor{boxcolor}{RGB}{193, 229, 255}
\definecolor{figboxcolor}{RGB}{193, 229, 255}
\definecolor{boxcolorgray}{RGB}{227, 227, 227}
\definecolor{goalcolor}{RGB}{229, 229, 97}
\definecolor{boxgoalcolor}{RGB}{229, 229, 97}
\definecolor{darkcyan}{rgb}{0.0, 0.0, 0.7}

\sethlcolor{boxcolorgray}

\def\gtrsim{\lower 2pt \hbox{$\, \buildrel {\scriptstyle >}\over
{\scriptstyle \sim}\,$}}
\def\lesssim{\lower 2pt \hbox{$\, \buildrel {\scriptstyle <}\over
{\scriptstyle \sim}\,$}}

\def\approxlt{\lower.2em\hbox{$\buildrel < \over \sim$}}
\def\approxgt{\lower.2em\hbox{$\buildrel > \over \sim$}}

\def\be{\begin{equation}}
\def\ee{\end{equation}}
\def\bea{\begin{eqnarray*}}
\def\eea{\end{eqnarray*}}

\def\jref@jnl#1{{\rm#1}}
\def\aj{\jref@jnl{AJ}}			
\def\araa{\jref@jnl{ARA\&A}}		
\def\apj{\jref@jnl{ApJ}}			
\def\apjl{\jref@jnl{ApJ}}		
\def\apjs{\jref@jnl{ApJS}}		
\def\ao{\jref@jnl{Appl.~Opt.}}		
\def\apss{\jref@jnl{Ap\&SS}}		
\def\aap{\jref@jnl{A\&A}}		
\def\aapr{\jref@jnl{A\&A~Rev.}}		
\def\aaps{\jref@jnl{A\&AS}}		
\def\azh{\jref@jnl{AZh}}			
\def\baas{\jref@jnl{BAAS}}		
\def\jrasc{\jref@jnl{JRASC}}		
\def\memras{\jref@jnl{MmRAS}}		
\def\mnras{\jref@jnl{MNRAS}}		
\def\pra{\jref@jnl{Phys.~Rev.~A}}	
\def\prb{\jref@jnl{Phys.~Rev.~B}}	
\def\prc{\jref@jnl{Phys.~Rev.~C}}	
\def\prd{\jref@jnl{Phys.~Rev.~D}}	
\def\pre{\jref@jnl{Phys.~Rev.~E}}	
\def\prl{\jref@jnl{Phys.~Rev.~Lett.}}	
\def\pasp{\jref@jnl{PASP}}		
\def\pasj{\jref@jnl{PASJ}}		
\def\qjras{\jref@jnl{QJRAS}}		
\def\skytel{\jref@jnl{S\&T}}		
\def\solphys{\jref@jnl{Sol.~Phys.}}	
\def\sovast{\jref@jnl{Soviet~Ast.}}	
\def\ssr{\jref@jnl{Space~Sci.~Rev.}}	
\def\zap{\jref@jnl{ZAp}}			
\def\nat{\jref@jnl{Nature}}		
\def\iaucirc{\jref@jnl{IAU~Circ.}}
\def\aplett{\jref@jnl{Astrophys.~Lett.}}
\def\apspr{\jref@jnl{Astrophys.~Space~Phys.~Res.}}
\def\bain{\jref@jnl{Bull.~Astron.~Inst.~Netherlands}}
\def\fcp{\jref@jnl{Fund.~Cosmic~Phys.}}
\def\gca{\jref@jnl{Geochim.~Cosmochim.~Acta}}
\def\grl{\jref@jnl{Geophys.~Res.~Lett.}}
\def\jcp{\jref@jnl{J.~Chem.~Phys.}}	
\def\jgr{\jref@jnl{J.~Geophys.~Res.}}	
\def\jqsrt{\jref@jnl{J.~Quant.~Spec.~Radiat.~Transf.}}
\def\memsai{\jref@jnl{Mem.~Soc.~Astron.~Italiana}}
\def\nphysa{\jref@jnl{Nucl.~Phys.~A}}
\def\physrep{\jref@jnl{Phys.~Rep.}}
\def\physscr{\jref@jnl{Phys.~Scr}}
\def\planss{\jref@jnl{Planet.~Space~Sci.}}
\def\procspie{\jref@jnl{Proc.~SPIE}}

\def\fas{\hbox{$.\!\!^{\prime\prime}$}}

\def\ion#1#2{#1$\;${\small\rm\@Roman{#2}}\relax}

\newcount\lecurrentfam
\def\LaTeX{\lecurrentfam=\the\fam \leavevmode L\raise.42ex
\hbox{$\fam\lecurrentfam\scriptstyle\kern-.3em A$}\kern-.15em\TeX}
\def\sizrpt{
(\fontname\the\font): em=\the\fontdimen6\font, ex=\the\fontdimen5\font
\typeout{
(\fontname\the\font): em=\the\fontdimen6\font, ex=\the\fontdimen5\font
}}
\definecolor{dark}{rgb}{ 0.34, 0.09, 0.49}

\usepackage{fancyhdr,url}
\usepackage{transparent}
\usepackage{tablefootnote}
\usepackage{mathpazo}

\begin{document}

\noindent
\begin{raggedright}
{\Huge\bfseries Star Formation across the Galaxy  \\[0.5em]
\Large Resolving fragmentation, accretion and feedback throughout the entire Milky Way
}\end{raggedright}\\

\noindent
\'Alvaro S\'anchez-Monge$^{1,*}$, 
Henrik Beuther$^{2}$, 
Maryvonne Gerin$^{3}$, 
Fr\'ed\'erique Motte$^{4}$, 
Gary Fuller$^{5}$, \\
Maite Beltr\'an$^{6}$, 
Elias Brinks$^{7}$, 
Gemma Busquet$^{8}$, 
Laura Colzi$^{9}$, 
Audrey Coutens$^{10}$, \\
Timea Csengeri$^{11}$, 
Claudia Cyganowski$^{12}$, 
Maria Drozdovskaya$^{13}$, 
Manuel Fern\'andez-L\'opez$^{1}$, \\
Francesco Fontani$^{6}$, 
Roberto Galv\'an-Madrid$^{14}$, 
Adam Ginsburg$^{15}$, 
Josep Miquel Girart$^{1}$, \\
\'Alvaro Hacar$^{16}$, 
Katharine Johnston$^{17}$, 
Jes K.\ Jørgensen$^{18}$, 
Kee-Tae Kim$^{19}$, 
Pamela Klaassen$^{20}$, \\
Michael K\"uffmeier$^{18}$, 
Qiuyi Luo$^{21}$, 
S\'ebastien Maret$^{4}$, 
Sergio Mart\'in$^{22}$, 
Chiara Mininni$^{23}$, \\
Kaho Morii$^{24}$, 
Alice Nucara$^{23}$, 
Thomas Nony$^{11}$, 
Fernando Olguin$^{25}$, 
Mayra Osorio$^{26}$, 
Aina Palau$^{14}$, \\
Jaime E.\ Pineda$^{27}$, 
V\'ictor M.\ Rivilla$^{9}$, 
Patricio Sanhueza$^{21}$, 
Miriam G.\ Santa-Maria$^{28}$, \\
Amelia Stutz$^{29}$, 
Alessio Traficante$^{23}$, 
Fengwei Xu$^{2}$, 
Maxime Valeille-Manet$^{30}$, 
Qizhou Zhang$^{24}$, \\
Birka Zimmermann$^{31}$
\vspace{0.5em}
\newline
{
\footnotesize{
\noindent $^*$ asanchez@ice.csic.es \\
\noindent Affiliations are listed at the end of the paper
}
}
\vspace{0.5em}

\begin{tcolorbox}[boxsep=3pt,left=0pt,right=0pt,top=1pt,bottom=1pt,colframe=boxcolor,colback=boxcolor]
\justify
\textbf{Abstract:} 
Star formation is both a multi-scale and highly dynamic process. It is regulated by the combination of gravity, turbulence, feedback and magnetic fields, which connect more than six orders of magnitude in scale: from large molecular clouds (several 10s to 100s pc in size) down to protoplanetary disks (a few 10s to 100s of astronomical units in size, or $10^{-5}$ to $10^{-4}$~pc). \textbf{\textit{Resolving the processes of fragmentation, accretion, and feedback that result in the formation of stellar systems throughout all relevant scales, from clouds to disks, and across the entire range of different environments in the Milky Way}} is key to constructing a comprehensive understanding and a predictive model of the formation process of stars of all masses that populate not only our Galaxy but the wider Universe across cosmic times. 
With current astronomical facilities, we are only able to grasp the details of star formation in $<5$\% of Galactic star-forming regions, being heavily biased towards those close to us. Thus, characterizing this multi-scale phenomenon at the necessary sensitivity and resolution across all Galactic environments requires the next generation (sub)millimeter-wavelength facility: ALMA2040.
\end{tcolorbox}


\vspace{0em}

{
\begin{flushleft}
\textbf{Keywords:} \it star formation, interferometry, spectral lines, continuum, ISM: clouds, bubbles, outflows
\end{flushleft}
}
\vfill

\newpage{}

\vspace{-1.0cm}
\section{Scientific Context and Open Questions}

\vspace{-0.05cm}
Star formation is an essential process that permeates the entire Universe. Newly-formed stars play an important role in shaping the morphological, dynamical and chemical structure of their host galaxies. Furthermore, these young stars also exert a major influence on the formation and development of planetary systems, which form along with their hosting stars.

\begin{figure}[t]
\centering
\includegraphics[width=0.99\linewidth]{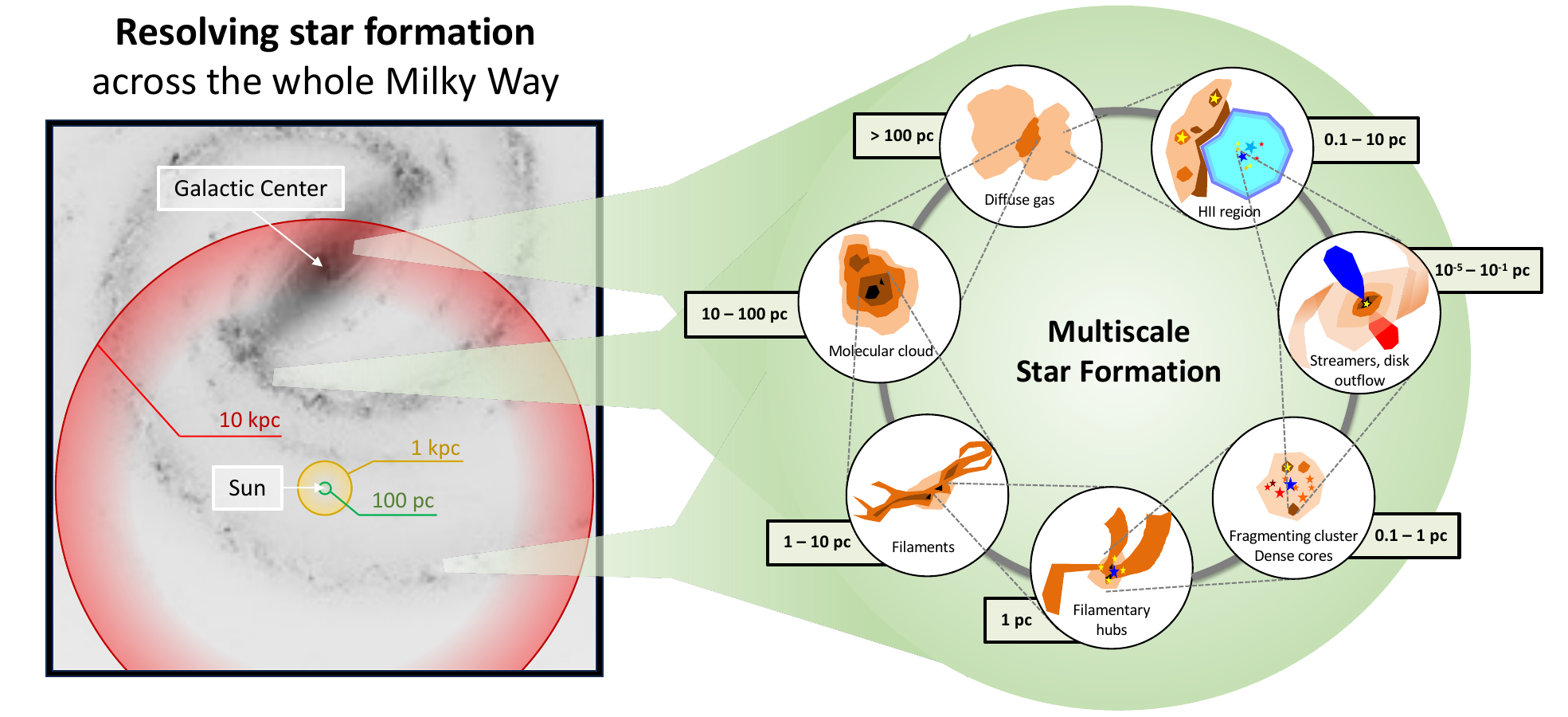}
\vspace{-0.15cm}
\caption{\textit{Sketch illustrating the science case of this white paper: ALMA2040 will resolve star formation, from large-scale clouds down to protostellar disks (as shown in the right-panel), in star-forming complexes distributed across the various environments of the Milky Way up to 10~kpc (as shown in the left panel).}
\vspace{-0.cm}}
\label{fig:visual}
\end{figure}

Despite its relevance, the multifaceted and multiscale nature of star formation renders it a complex scientific problem. In essence (see Fig.~\ref{fig:visual} right panel), stars form inside large molecular clouds (several 10 to 100~pc in size, with masses up to $10^6$~$M_\odot$) that divide into a series of filaments (a few sub-pc in extent) down to individual magnetized dense gas cores undergoing gravitational collapse and slow rotation (about 0.01--0.1~pc in size), where accretion disks (a few 10s to 100s~au in size, with masses $\sim$~$M_\odot$) channel material toward the forming star \citep[e.g.,][]{Hacar2023, Pineda2023, Beuther2025}. Thus, a comprehensive understanding of star formation necessitates the holistic study of multiple physical processes (e.g., gravity, turbulence, magnetic fields) and their interactions through many orders of magnitude in spatial and mass scales.

In the last decade, thanks to facilities like ALMA, significant steps forward have been made in our understanding of star formation. For instance, the structure of protostellar disks has been resolved in nearby sources \cite[e.g.,][]{Oberg2021, Ohashi2023}; accreting streamers feeding disks and stars have been mapped \citep[e.g.,][]{Alves2019, Gieser2025, GallowaySprietsma2026}; the role of gravity, turbulence and magnetic fields has been studied in a few young stellar objects \citep[YSOs, e.g.,][]{Beltran2019, Huang2024, Liu2026}; disks and jets have been resolved around a few high-mass protostars \citep[e.g.,][]{Ginsburg2018, Cesaroni2025, SanchezMonge2025}; and studies of massive protoclusters have revealed that dynamical processes are crucial to the origin of stellar masses \citep[e.g.,][]{Motte2018b, Xu2024}.

Unfortunately, \textbf{resolving all the relevant spatial scales} with the resolution and sensitivity required to characterize the fragmentation process, the mass transport, and the impact of feedback is \textbf{currently only possible in nearby star-forming regions within the solar neighborhood} (e.g., out to 100s~pc distances, like the Taurus or Orion star-forming complexes), \textbf{or towards some of the brightest star-forming regions} in the Galaxy. \textit{These exceptionally bright or nearby regions are unlikely to be representative of the regions where stars in our Galaxy typically form}. Despite pushing current astronomical facilities to their limits (in sensitivity, resolution, and wide field coverage; in many cases trading off one or two of these limits), we are only able to grasp the details of star formation in $<5$\% of Galactic star-forming regions, and we are heavily biased towards those close to us.

\vspace{-0.1cm}
\begin{tcolorbox}[boxsep=3pt,left=0pt,right=0pt,top=1pt,bottom=1pt,colframe=boxgoalcolor,colback=goalcolor]
To significantly advance our understanding of star formation and provide answers to several outstanding questions (see Sect.~\ref{sec:goals}), we need \textbf{to resolve the multi-scale processes that regulate star formation in different environments distributed across the entire Milky Way} (see Fig.~\ref{fig:visual}), encompassing environments as diverse as the Galactic center, spiral arms and the outer Galaxy.
\end{tcolorbox}
\vspace{-0.1cm}

While star formation is a ubiquitous process throughout the Universe, it is only in our own galaxy, the Milky Way, where we can access all the relevant scales, from clouds to disks. Moreover, the existence of different Galactic environments ranging from the chemically-enriched, highly-energized Central Molecular Zone around the Galactic center, to the typical spiral arm star-forming complexes, and up to the metal-poor outer Galaxy, permits us to concretely assess the impact of varied physical conditions on the physics of star formation, bridging these studies to distant extragalactic systems where resolution and sensitivity do not permit the dissection of individual star forming objects. In summary, resolving star formation in different Galactic environments is the key to understanding how typical is the star formation that we have studied in the solar neighborhood and to deciphering star formation across cosmic time throughout the Universe. 

%
\vspace{-0cm}
\section{Science Cases for the 2040's\label{sec:goals}}

\vspace{-0.05cm}
To make significant progress in understanding star formation in our Galaxy requires capabilities significantly greater than available from current generation facilities. To this end we identify four key science cases that, with next generation capabilities, would transform our understanding of star formation, one of the most fundamental processes in the evolution of galaxies and the Universe itself. These science cases address the levels of {\textbf{fragmentation and multiplicity}} within star-forming regions across the Galaxy (see Sect.~\ref{sec:fragmentation}), the {\textbf{hierarchical mass and energy flow}} from large-cloud scales down to disks and stars (see Sect.~\ref{sec:accretion}), the influence of {\textbf{mechanical and radiative feedback}} from newly formed stars (see Sect.~\ref{sec:feedback}), and the {\textbf{formation and early evolution of the most massive stars}} (see Sect.~\ref{sec:highmass}). Providing answers to these science cases requires an innovative and transformative ground-based millimeter-wavelength facility with high resolving power, high sensitivity and high image fidelity, as well as efficient mapping of large fields of views, i.e., \textbf{{these science cases necessitate of ALMA2040}} (see Sect.~\ref{sec:technical} for the technical details).

\vspace{-0cm}
\subsection{Determining the initial stellar population: Fragmentation and multiplicity \label{sec:fragmentation}}

\vspace{-0.05cm}
The mechanisms responsible for the formation of stellar clusters and associations, and by extension the origin of the initial mass function \citep[IMF;][]{Salpeter1955, Kroupa2026} of stars, remains one of the long-standing unsolved problems in the field of star formation \citep[e.g.,][]{Offner2014}. Directly related to the IMF is the understanding of the formation mechanisms of stellar multiple systems (e.g., core fragmentation during early collapse, disk fragmentation, dynamical capture; \citealt[][]{Offner2023}), and how these different proposed modes may depend on the environment.

Despite recent attempts to study the fragmentation process and the multiplicity levels in star-forming regions  \citep[see e.g.,][]{Palau2018, Sanhueza2019, Li2024, Coletta2025}, we are still far from the required spatial resolution, sensitivity and statistical robustness to significantly improve our understanding of the fragmentation process across Galactic environments. For instance, the ALMA large programs ALMA-IMF \citep[conducting mosaic-observations of 15 star-forming regions;][]{Motte2022}, ALMAGAL \citep[targeting small portions of 1000+ star-forming regions across the Galaxy;][]{Molinari2025} and ACES \citep[surveying the Central Molecular Zone around the Galactic center;][]{Longmore2026} are able to study fragmentation only down to spatial scales of several 1000s~au and down to mass completeness levels of $\approx0.3$~$M_\odot$ (i.e.\ far from the required $\approx50$~au and $\approx0.01$~$M_\odot$ levels, see below). Higher spatial-resolution studies are only feasible towards reduced samples of objects, as shown e.g., by the DIHCA project \citep{Olguin2021}, where 23 objects embedded within large stellar (proto)clusters have been observed at a resolution of $\approx160$~au with mass sensitivities of $\approx0.1$~$M_\odot$, providing evidence of core fragmentation as a significant mode for the formation of multiples in this kind of environments, but lacking resolution and sensitivity to probe the other formation modes \citep[][]{Luo2026}. Disk fragmentation is predicted as the main formation path candidate for tightly bound multiple systems \citep[e.g.,][]{Kratter2016}. Hence, resolving circumstellar disks ($\approx50$~au scales) across the full range of stellar masses will be crucial for understanding stellar multiplicity, and by that also the IMF.

\begin{figure}[t!]
\centering
\includegraphics[width=\linewidth]{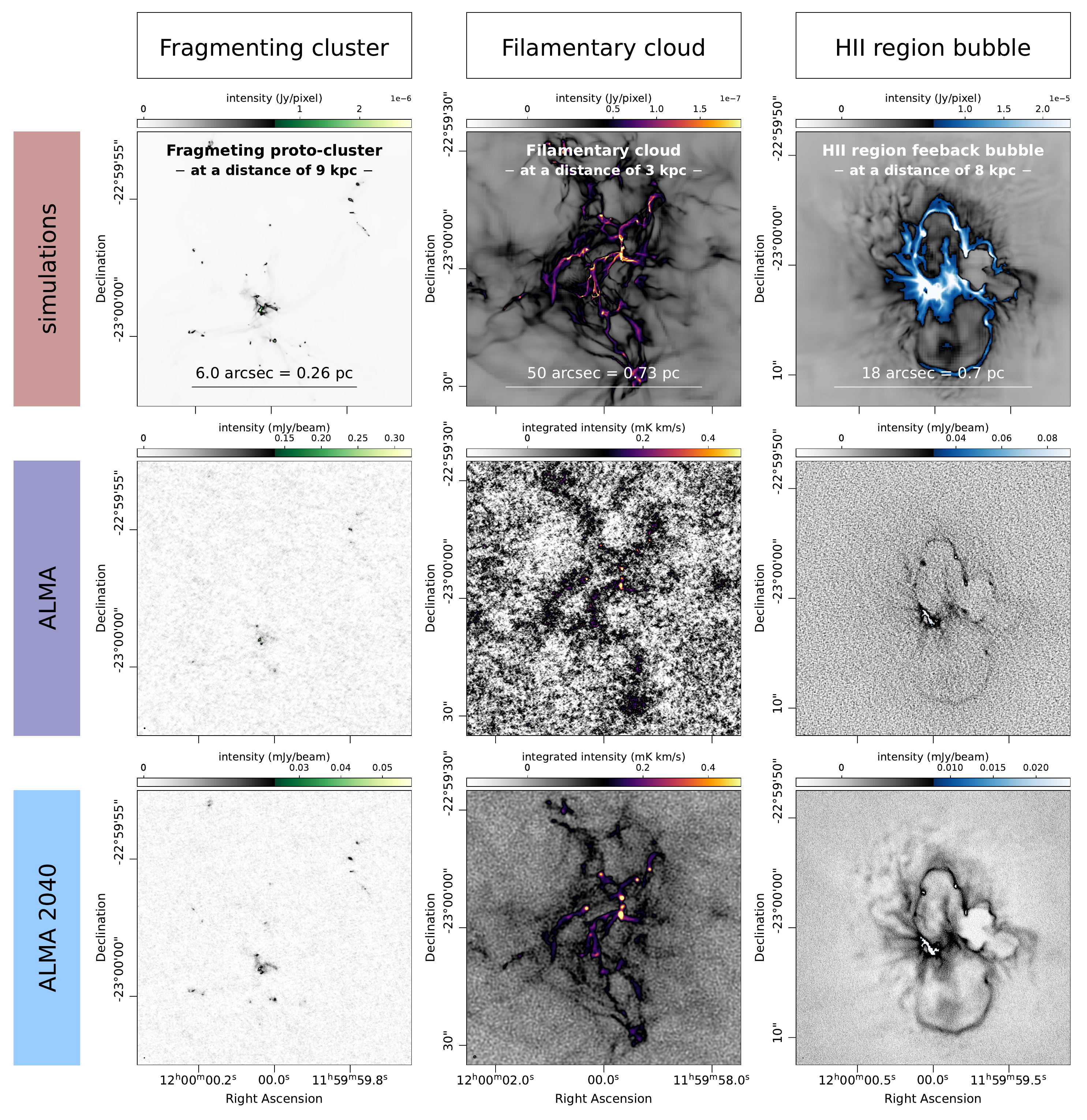}
\caption{\textit{Numerical simulations (\textit{top row}) and synthetic observations maps (\textit{middle} and \textit{bottom rows}) of (\textit{left column}) the 230~GHz continuum emission of a \textbf{fragmenting proto-cluster} with massive cores in the center and low/intermediate-mass cores distributed all over the field, (\textit{central column}) the equivalent C$^{18}$O\,(2--1) integrated line emission of a \textbf{filamentary cloud} with several dense hubs, and (\textit{right column}) the 230~GHz continuum emission of an \textbf{HII region feedback bubble} driven by massive stars forming at the center of the cloud. The \textit{top row} shows the emission maps at 230~GHz produced with RADMC-3D from the numerical simulations of \citet{Zimmermann2025, Zimmermann2026}. Synthetic images using state-of-the-art ALMA configurations are shown in the \textit{middle row} panels, while the results from a next generation millimeter facility (ALMA2040, see Sect.~\ref{sec:technical}) are shown in the \textit{bottom row}. Table~\ref{tab:simulations} presents a quantitative comparison (configurations used, resolution, sensitivity, and dynamic range) of the ALMA and ALMA2040 images.}}
\label{fig:simulations}
\end{figure}

To address the fragmentation and multiplicity levels across Galactic environments, we anticipate a next generation millimeter facility that, on the one hand, can resolve very small spatial structures ($\approx50$~au at 10~kpc distance) at the high continuum and line sensitivity necessary to investigate the fragmentation down to the brown dwarf limit ($\approx0.01$~$M_\odot$), and that at the same time covers large enough areas ($10\times10$~pc$^{2}$) to investigate the cluster-forming regions within their environments. Only with high sensitivity and spatial resolution, will we be able to constrain the core mass function and its potential relation to the IMF down to the brown dwarf limit \citep[e.g.,][see ALMA2040 white paper by Palau et al]{Palau2024}, and thereby constrain the formation of stars and multiples over the entire mass range. The left-column panels of Fig.~\ref{fig:simulations} show the outcome of a numerical simulation of a fragmenting (proto)cluster at a distance of 9~kpc, and how ALMA2040 observations will resolve and identify all relevant cores and fragments, compared to the limited view of ALMA observations that are only sensitive to the most massive objects and lack resolving power to distinguish close binaries. \textit{\textbf{Thanks to ALMA2040, we aim at resolving and studying the fragmentation and multiplicity levels of any star-forming complex distributed throughout the Galaxy, from the Central Molecular Zone around the Galactic center up to the outer Galaxy}}.

\begin{table}[t]
\caption{Parameter comparison of the ALMA and ALMA2040 synthetic images shown in Fig.~\ref{fig:simulations}}
\label{tab:simulations}
\small
\vspace{-0.2cm}
\begin{tabular}{p{0.40\textwidth} p{0.25\textwidth} p{0.25\textwidth}}
\hline
\noalign {\smallskip}
  \textbf{Science case}
& \multicolumn{1}{c}{\textbf{ALMA}}
& \multicolumn{1}{c}{\textbf{ALMA2040}}
\\
\hline
\noalign {\smallskip}
\multicolumn{1}{l}{\textbf{Fragmenting protocluster at 9~kpc distance}} \\
--- array configuration                & \multicolumn{1}{c}{C--9 + C--6}              & \multicolumn{1}{c}{Extended} \\
--- total on-source time$^\dagger$ (h) & \multicolumn{1}{c}{1.0}                      & \multicolumn{1}{c}{1.0} \\
--- angular resolution (arcsec)        & \multicolumn{1}{c}{$0\fas037\times0\fas031$} & \multicolumn{1}{c}{$0\fas015\times0\fas015$} \\
--- rms noise level ($\mu$Jy/beam)     & \multicolumn{1}{c}{17}                       & \multicolumn{1}{c}{4} \\
--- dynamic range low-mass cores       & \multicolumn{1}{c}{$\approx$2}               & \multicolumn{1}{c}{15} \\
\hline
\noalign {\smallskip}
\multicolumn{1}{l}{\textbf{Filamentary cloud at 3~kpc distance}} \\
--- array configuration                & \multicolumn{1}{c}{C--5 + C--2 + ACA + TP}   & \multicolumn{1}{c}{Compact + TP} \\
--- total on-source time$^\dagger$ (h) & \multicolumn{1}{c}{2.0}                      & \multicolumn{1}{c}{2.0} \\
--- angular resolution (arcsec)        & \multicolumn{1}{c}{$0\fas65\times0\fas63$}   & \multicolumn{1}{c}{$0\fas65\times0\fas63$} \\
--- rms noise level (mK km/s)          & \multicolumn{1}{c}{105}                      & \multicolumn{1}{c}{11} \\
--- dynamic range$^\ddagger$ filaments & \multicolumn{1}{c}{$\approx$5}               & \multicolumn{1}{c}{56} \\
\hline
\noalign {\smallskip}
\multicolumn{1}{l}{\textbf{HII region feedback bubble at 8~kpc distance}} \\
--- array configuration                & \multicolumn{1}{c}{C--9 + C--6 + C--3 + ACA} & \multicolumn{1}{c}{Extended + Intermediate} \\
--- total on-source time$^\dagger$ (h) & \multicolumn{1}{c}{2.0}                      & \multicolumn{1}{c}{2.0} \\
--- angular resolution (arcsec)        & \multicolumn{1}{c}{$0\fas038\times0\fas032$} & \multicolumn{1}{c}{$0\fas024\times0\fas022$} \\
--- rms noise level ($\mu$Jy/beam)     & \multicolumn{1}{c}{15}                       & \multicolumn{1}{c}{4} \\
--- dynamic range extended emission    & \multicolumn{1}{c}{$\approx$3}               & \multicolumn{1}{c}{20} \\
\hline
\multicolumn{3}{c}{\small{$^\dagger$ Total on-source time after counting the on-source time spent in each array configuration.}} \\
\multicolumn{3}{c}{\small{$^\dagger$ The dynamic range of the entire ALMA2040 image, i.e.\ considering the bright compact objects, is $\approx600$.}} \\
\end{tabular}
\vspace{-0.15cm}
\end{table}

\vspace{-0cm}
\subsection{Hierarchical mass and energy flow: From cloud to stellar masses\label{sec:accretion}}

\vspace{-0.05cm}
The star formation process is closely linked to the cascade of energies and gas flows that develop across multiple scales, from pc-size molecular clouds down to sub-pc dense cores collapsing into disks and stars. In the past decade, surveys of nearby star-forming regions have revealed the clouds' multifractal structure: consisting of filamentary networks, clumps, cores and disks \citep[e.g.,][]{Hacar2023, Pineda2023}. These works show that hierarchical gas flows connect these different scales, with directionality (inflow or outflow) determined by the evolutionary stage.

Completely characterizing gas flows and quantifying their associated energy and mass budgets is a significant challenge that current facilities can only ever partially address \citep[e.g.,][]{TrevinoMorales2019, FernandezLopez2023, Olguin2025}. The first difficulty in such studies arises from the need to select multiple molecular line tracers to track these gas flows (with velocity gradients $\approx1$~km~s$^{-1}$~pc$^{-1}$) across different spatial (from a few pc down to $\approx10^{-3}$~pc) and density (from $10^3$~cm$^{-3}$ to $10^8$~cm$^{-3}$) scales. For instance, while molecular species such as CO isotopologues and N$_2$H$^+$ are sensitive to large-scale filamentary structures \citep[e.g.,][]{Hacar2018, Suri2019}, the line emission of these species becomes optically thick at smaller and denser scales, requiring to select other molecular tracers sensitive to higher critical densities. Species such as C$^{18}$O, HC$_3$N, H$_2$CO, and CS are then necessary to reveal accretion streamers in the vicinity of protostellar disks around low-mass YSOs \citep[e.g.,][]{Pineda2020, ValdiviaMena2022}. For regions with on average higher densities (e.g., massive YSOs), observations of complex molecular species with higher critical densities (e.g.,  CH$_3$OH, CH$_3$OCHO) are required \citep[e.g.,][]{Schwoerer2019, Olguin2025}. Multiple molecular line tracers are thus needed to detect and characterize the broad range of spatial and density scales associated with the mass flow from clouds to stars. The second difficulty concerns the need to perform high-resolution multi-scale studies in a variety of environments (in terms of density, irradiation, shocks, metallicity; see ALMA2040 white paper by Fontani et al.) that must be studied to understand star formation throughout the Milky Way. Current facilities are not sensitive enough to spectral lines tracing a broad range of densities to enable kinematic studies across multiple spatial scales in star-forming complexes located beyond the solar neighborhood or beyond a few bright star-forming complexes in the Galaxy.

To characterize the gas flow across a broad range of spatial and density scales throughout the different Galactic environments requires very large, spectrally-resolved ($\approx0.1$~km~s$^{-1}$) multi-line surveys and high-fidelity images with high dynamic ranges ($>1\,000$), covering filaments ($\approx1$~pc), streamers ($\approx0.05$~pc) and disks ($\approx50$~au). This can be achieved with a next generation millimeter facility that simultaneously provides high angular resolution and high image fidelity beyond the current ALMA capabilities. The central-column panels of Fig.~\ref{fig:simulations} show a numerical simulation of a filamentary cloud and how the high-fidelity imaging of ALMA2040 recovers the whole filamentary structure, while ALMA observations can only partially guess the presence of these elongated structures. \textit{In summary, \textbf{ALMA2040 is essential to map the cascade of mass and energy flows across spatial and density scales in the full range of Galactic environments, in star-forming regions located at multiple kpc distance within the Milky Way}}.

\subsection{Regenerating the interstellar medium: Mechanical and radiative feedback\label{sec:feedback}}

\vspace{-0.05cm}
The feedback from high-energy radiation, energetic particles and through mechanical actions is key in defining the structure of interstellar clouds and their ability to form stars. These mechanisms operate on all scales, from molecules and dust grains to cloud interfaces, potentially creating sharp gradients in all relevant physical and chemical quantities \citep[see e.g.,][]{Bally2016, Schneider2020, Grudic2021, Neralwar2024, Zimmermann2025}.

In star-forming regions, feedback may drive successive generations of stars, altering the structure, chemistry and ionization fraction of prestellar cores, protostellar envelopes and disks \citep[e.g.,][]{Masque2009, Schneider2020, Cosentino2026}, and potentially the mass of the newly-formed stars. For example, UV radiation emitted by massive stars leads to the photo-evaporation of nearby protoplanetary disks (proplyds), modifying the amount of mass available to the star and its surrounding planets. Recently, \citet{Boyden2025} have successfully detected radio recombination lines (RRLs) at mm wavelengths towards the proplyds in Orion, paving the way for investigating feedback effects in these systems.
In general, spectral line diagnostics at (sub)mm wavelengths (e.g., radiation-sensitive species: CCH, CN, CO$^{+}$, RRLs; shock tracers: SiO, CH$_3$OH) --- complementing studies in the infrared and radio with present and future facilities such as JWST, ELT, PRIMA, SKA and ngVLA --- are essential to assess the magnitude and efficiency of these feedback processes. These line diagnostics, that can only be obtained by accessing the line profile of key species and transitions across the (sub)mm spectral range, provide quantitative ways to evaluate the impact of feedback on local and global scales (e.g., gas composition, thermal balance, dynamics) when compared with sophisticated models \citep[e.g.,][]{Grudic2021, Pound2023, Palud2025, Neralwar2025}.
Although state of the art models and simulations can be efficiently used to extract the main physical parameters from relevant spectral maps, the quality and sensitivity of current line observations do not match that of the models. Current astronomical facilities can only obtain the required information on either small areas or relatively small selected spectral ranges.

A leap forward in line sensitivity (10x better than current facilities), together with improved image fidelity, is therefore needed to allow the exploration of feedback mechanisms over a broad range of core and cloud scales. The right-column panels of Fig.~\ref{fig:simulations} show an H\textsc{ii}-region feedback bubble located at 8~kpc (20$\times$ the distance to Orion). ALMA observations are only sensitive to a fraction of the brightest and sharpest edges of the bubble, while ALMA2040 observations recover the whole extend of the structure, including the more diffuse component. At smaller scales, this leap in line sensitivity will enable the detection of RRL emission from proplyds in star-forming regions beyond Orion (e.g., M17, NGC\,6334, at kpc distances) increasing the range of environments where feedback processes can be studied.
Finally, although stars of all masses exert feedback, past studies only had the sensitivity to focus on the impact of the most massive stars (which are more energetic but less numerous) in selected star-forming regions, thus limiting an accurate understanding of the energy injection (e.g., outflows, jets, winds) by young stars of all masses in their surrounding medium across different environments. \textit{In summary, \textbf{ALMA2040 will allow to measure and establish the impact that stars of different masses exert on their natal environment, and how this affects the star formation process in different habitats across the Milky Way}}.

\subsection{Forging giants: Formation end early evolution of the most massive stars\label{sec:highmass}}

\vspace{-0.05cm}
Among the various stars that form in star-forming complexes, the most massive ones (those exceeding 8~$M_\odot$) are the least well understood. These stars are rare in number (only 2 to 3 for every 1000 solar-type star) and have very short lifetimes (only a few million years), but they have luminosities thousands of times greater than solar-type stars and drive energetic phenomena that impact and shape their surroundings as well as their host galaxies.

Current models of massive star formation suggest that accretion via disks ($\approx100$ au in radius) and large-scale envelopes ($>$1000~au) are likely responsible for the growth of high-mass protostars \citep[see e.g.,][]{Kuiper2010, Motte2018a, Beuther2025}. Unfortunately, testing these models is challenging because of current observational limitations. Massive stars typically form at large distances, within densely crowded and highly obscured regions, that require observations with high spatial resolution and sensitivity at (sub)mm wavelengths. Facilities like ALMA have greatly enhanced our knowledge \citep[see e.g.,][and references below]{Hunter2026, Wang2026}; however, these studies often focus on a few objects and fall short of achieving the necessary spatial resolution, maximum recoverable scale, sensitivity, statistical robustness (in terms of the number of sources) and sampling of different galactic environments.

The most pressing questions regarding the formation of massive stars are: (i)~The mechanisms of accretion and mass assembly through filaments, streamers and disks \citep[e.g.,][]{AnezLopez2020, Sanna2025, Olguin2026}, as well as the timing and manner in which radiative feedback begins and its actual influence on the continued accretion \citep[e.g., ionized accretion flows,][]{Keto2008, GalvanMadrid2023}; (ii)~The structure, stability and dynamics of disks around massive protostars \citep[e.g.,][]{Jankovic2019, Ahmadi2023}, which might be studied through sensitive observations of (semi)refractory compounds \citep[e.g., NaCl, KCl;][]{Ginsburg2019, Ginsburg2023} enabling access to the innermost regions of these massive and hot disks; and (iii)~The formation process of the most massive and extreme objects, along with its potential connection to the presence of explosive outflows \citep[see review][]{Zapata2026}. These explosive outflows --- different from classical bipolar outflows associated with accretion events --- show dozens of collimated ejections in a radial and almost isotropic fashion. Over the last decade, sensitive ALMA spectral line observations have revealed the existence of a few explosive outflows \citep[e.g.,][]{Bally2017, GuzmanCcolque2024}, and the most recent estimates establish a rate of events comparable to that of supernovae and massive star formation (about one every $\sim$100~years). 

To provide a comprehensive picture of the formation process of the most massive stars requires resolving $\approx50$~au scales at several kpc distance, with enough image fidelity and line sensitivity (5--10$\times$ improvement with respect to ALMA/WSU). The greater spatial resolution and sensitivity will resolve the disks' structure (see ALMA2040 white paper by K\"uffmeier et al.), identify the presence of ionized accretion flows, and enable the study of their kinematics via spectral lines of e.g., (semi)refractory compounds (see ALMA2040 white paper by Rivilla et al.). The enhanced capabilities will also allow to compile a complete census of all explosive outflows at distances $>5$~kpc, and to identify and characterize the objects that originate them, with the possibility of measuring proper motions \citep[$\approx10$~mas~yr$^{-1}$,][]{Rodriguez2017} over timescales of a few years.
\textit{In summary, \textbf{addressing these pressing issues and gaining deeper insight into the fragmentation, accretion and feedback processes related to the formation and early evolution of high-mass stars necessitates resolving the gas and dust surrounding massive protostars with high spatial resolution and enhanced line sensitivity}}. \textit{The next generation millimeter facility ALMA2040 will constitute a significant breakthrough in the study of high-mass star formation across the whole Galaxy, akin to how ALMA has revolutionized our understanding of local star formation since its 2014 Long Baseline Campaign}.

\begin{figure}[t]
\centering
\includegraphics[width=0.94\linewidth]{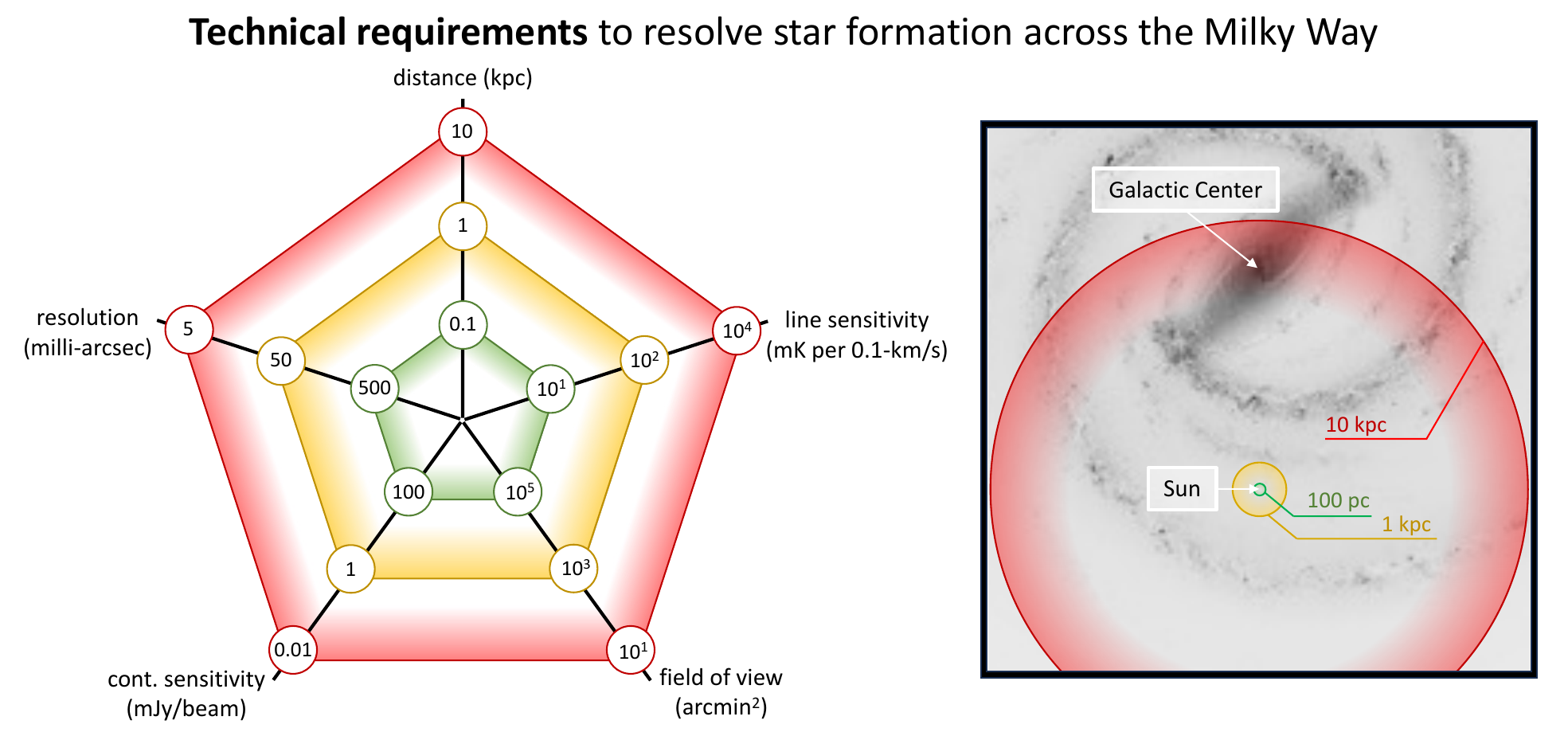}
\vspace{-0.2cm}
\caption{\textit{(\textbf{left panel}): Angular resolution, continuum sensitivity, field of view, and line sensitivity necessary to resolve a 50-au size object, with a gas+dust mass down to 0.01~$M_\odot$, over regions of 10$\times$10~pc$^{2}$, with a line sensitivity necessary to detect filaments, disks and outflows, throughout the Galaxy at distances of 100~pc (green), 1~kpc (yellow) and 10~kpc (red). (\textbf{right panel}): Face-on view of the Milky Way, highlighting the distances at which fragmentation, accretion and feedback can be studied with the corresponding requirements (left panel).} \textbf{Current ALMA observations are able to partially resolve star-forming regions up to 100--1000~pc} \textit{(green and yellow areas, i.e.\ local neighborhood).} \textbf{ALMA2040 will probe regions up to 10~kpc distances} \textit{(red area, i.e.\ through the full range of Galactic environments).}
\vspace{-0.1cm}}
\label{fig:sketch}
\end{figure}

%
\vspace{-0.4em}
\section{ALMA2040: Technical Requirements\label{sec:technical}}

\vspace{-0.2em}
Figure~\ref{fig:sketch} outlines the technical specifications necessary to achieve the scientific goals of Sect.~\ref{sec:goals}. Specifically, these goals necessitate \textbf{ALMA2040}, a next-generation (sub)millimeter interferometer with the following requirements: \textbf{(1)}~a $\ge$10x increase in instantaneous \textbf{spectral line sensitivity} in Bands~2--10 (70--950~GHz), compared to current ALMA ($\approx$8x compared to anticipated ALMA-WSU), and a 10x increase in continuum sensitivity, compared to current ALMA ($\approx$5x compared to anticipated ALMA-WSU); \textbf{(2)}~a $\approx$10x smaller PSF area through 
\textbf{3--5x longer baselines} than ALMA-WSU; \textbf{(3)}~access to large angular scales that could be achieved by combination with total-power observations, ensuring 
\textbf{enhanced image fidelity} for regions with emission at multiple spatial scales; and \textbf{(4)}~\textbf{fast mapping speeds} (e.g., thanks to the enhanced sensitivity) to efficiently map large mosaic areas, providing easy access to larger fields of views ($\approx$100x than current ALMA). All these requirements can be achieved primarily by increasing the collecting area compared to ALMA-WSU and by disruptive developments in receiver technology, while at the same time providing exquisite image fidelity (through exquisite uv-coverage) across a range of angular scales.

Figure~\ref{fig:simulations} (and Table~\ref{tab:simulations}) shows the potential of ALMA2040 compared to state-of-the-art ALMA observations. ALMA2040 is conceived as a 200--300 antennas array. The ALMA2040 synthetic observations shown in Fig.~\ref{fig:simulations} have been generated using the ALMA GHOST tool\footnote{Developed and hosted by the UK ALMA Regional Centre Note, see \url{https://alma-ghost.jb.man.ac.uk/}}, which considers a 255$\times$12m array with three configurations: compact ($\approx$0.4~km max.\ baseline), intermediate ($\approx$12.5~km max.\ baseline) and extended ($\approx$28~km max.\ baseline), plus a Total Power (TP) array with 5$\times$24m telescopes. The large number of antennas not only leads to a significant enhancement in sensitivity but also offers a marked improvement in image fidelity due to the much better uv-coverage. Access to an efficient TP array is crucial for achieveing the ultimate level of image quality (see Fig.~\ref{fig:simulations} central column).
In summary, \textbf{\textit{this next-generation facility --- ALMA2040 --- will provide the required resolution, sensitivity and image fidelity, to detect and study star-forming complexes like the nearby Taurus/Orion-regions at kpc-distances throughout the Milky Way.}}

%
\vspace{-0.6em}
\section{Synergies with current and future facilities}

\vspace{-0.2em}
The \textbf{ALMA2040} science goals presented above will greatly benefit from synergies with current and future facilities sensitive to other regimes of the electromagnetic spectrum. For instance, the next generation of cm-wavelength interferometers, \textbf{SKAO} and \textbf{ngVLA}, will provide information about the thermal and non-thermal radio emission originated by HII regions, radio jets and accreting disks around YSOs, thus complementing the picture of the dust and gas emission observable primarily with ALMA2040. Current and next generation near/mid infrared telescopes, \textbf{JWST} and \textbf{ELT}, will give access to the embedded population of (proto)stellar objects, the emission from hot dust around them, as well as key spectral lines sensitive to ejection and accretion events. In the far-infrared, the space mission \textbf{PRIMA} (selected for Phase B implementation) will provide sensitive observations of the cold dust and magnetic fields at cloud scales. Finally, distances and proper motions of embedded YSOs can be studied with \textbf{Gaia} for nearby (embedded) star-forming regions (up to $\approx$1~kpc), and with the future \textbf{GaiaNIR} for more distant and obscured star forming complexes distributed throughout the Galaxy.

\section*{Acknowledgements}

\vspace{-0.1cm}
This work was developed as part of the broader ALMA2040 community effort. We acknowledge the Lorentz Center and NOVA for supporting the community workshops and coordination activities that helped initiate and shape the ALMA2040 White Paper process.
A.S.-M.\ acknowledges support from the PID2023-146675NB-I00 grant funded by MCIN/AEI/10.13039/501100011033 and by the European Union `Next GenerationEU’/PRTR. \\

\bibliographystyle{apj_mod}
{\bibliography{bibliography}}

\vspace{-0.0em}
\medskip
\noindent\makebox[\linewidth]{\rule{\textwidth}{0.4pt}}

{
\footnotesize{
\noindent 
$^{1}$~Institut de Ci\`encies de l’Espai (ICE-CSIC), Carrer de can Magrans s/n, 08193 Bellaterra, Barcelona, Spain \\
$^{2}$~Max-Planck-Institut f\"ur Astronomie, Königstuhl 17, 69117 Heidelberg, Germany \\
$^{3}$~LUX, Observatoire de Paris, PSL Research University, CNRS, Sorbonne Universit\'e, 75014 Paris, France \\
$^{4}$~Univ.\ Grenoble Alpes, CNRS, IPAG, 38000 Grenoble, France \\
$^{5}$~Jodrell Bank Centre for Astrophysics, The University of Manchester, Oxford Road, Manchester M13 9PL, UK \\
$^{6}$~INAF - Osservatorio Astrofisico di Arcetri, Largo E.\ Fermi 5, 50125 Firenze, Italy \\
$^{7}$~Centre for Astrophysics Research, University of Hertfordshire, College Lane, Hatfield AL10 9AB, UK \\
$^{8}$~Dpt.\ de F\'isica Qu\`antica i Astrof\'isica, Universitat de Barcelona (UB), Mart\'i i Franqu\`es 1, 08028 Barcelona, Spain \\
$^{9}$~Centro de Astrobiolog\'ia (CAB, CSIC-INTA), Ctra.\ de Ajalvir km.\ 4, Torrej\'on de Ardoz, 28850 Madrid, Spain \\
$^{10}$~Univ.\ Toulouse, CNES, CNRS, IRAP, Toulouse, France \\
$^{11}$~Laboratoire d’Astrophysique de Bordeaux, Univ.\ Bordeaux, CNRS, UMR 5804, 33615 Pessac, France \\
$^{12}$~School of Physics and Astronomy, University of St Andrews, North Haugh, St Andrews, Fife KY16 9SS, UK \\
$^{13}$~DCBP, Universit\"at Bern, Freiestrasse 3, 3012 Bern, Switzerland \\
$^{14}$~Instituto de Radioastronom\'ia y Astrof\'isica (IRyA-UNAM), Ant.\ Carretera a P\'atzcuaro 8701, 58089, Morelia, Mexico \\
$^{15}$~Department of Astronomy, University of Florida, PO Box 112055, Gainesville, FL, USA \\
$^{16}$~Institute for Astronomy (IfA), University of Vienna, Türkenschanzstrasse 17, A-1180 Vienna \\
$^{17}$~School of Engineering and Physical Sciences, University of Lincoln, Brayford Pool, Lincoln LN6 7TS, UK \\
$^{18}$~Niels Bohr Institute, University of Copenhagen, Jagtvej 155A, 2200 Copenhagen N, Denmark \\
$^{19}$~Korea Astronomy and Space Science Institute, 776 Daedeokdae-ro, Yuseong-gu, Daejeon 34055, Republic of Korea \\
$^{20}$~UK Astronomy Technology Centre, Royal Observatory Edinburgh, Blackford Hill, Edinburgh EH9 3HJ, UK \\
$^{21}$~Department of Astronomy, School of Science, The University of Tokyo, 7-3-1 Hongo, Bunkyo, Tokyo 113-0033, Japan \\
$^{22}$~European Southern Observatory, Alonso de C\'ordova 3107, Vitacura 763 0355, Santiago, Chile \\
$^{23}$~INAF-Istituto di Astrofisica e Planetologia Spaziale, Via Fosso del Cavaliere 100, 00133 Roma, Italy \\
$^{24}$~Center for Astrophysics | Harvard \& Smithsonian, 60 Garden Street, Cambridge, MA 02138, USA \\
$^{25}$~Kyoto University, Kitashirakawa Oiwakecho, Sakyo-ku, Kyoto 606-8502, Japan \\
$^{26}$Instituto de Astrofísica de Andalucía, CSIC, Glorieta de la Astronomía s/n, 18008 Granada, Spain \\
$^{27}$~Max-Planck-Institut für Extraterrestrische Physik, Giessenbachstraße 1, 85748 Garching, Germany \\
$^{28}$~Instituto de F\'isica Fundamental (CSIC), Calle Serrano 121-123, 28006 Madrid, Spain \\
$^{29}$~Departamento de Astronom\'ia, Universidad de Concepci\'on, Casilla 160-C, 4030000 Concepci\'on, Chile \\
$^{30}$~School of Physics and Astronomy, Cardiff University, Cardiff, CF24 3AA, UK \\
$^{31}$~I.\ Physikalisches Institut, Universit\"at zu K\"oln, Z\"ulpicher Str.\ 77, 50937 K\"oln, Germany
}
}

\end{document}